\documentclass[aps,prl,groupedaddress,twocolumn,10pt, reprint,superscriptaddress, longbibliography]{revtex4-2}

\usepackage{amsmath}
\usepackage{amssymb}
\usepackage{amsthm}
\usepackage{bbm}
\usepackage{graphicx}
\usepackage{amsmath}
\usepackage{graphicx}
\usepackage{color}
\usepackage{upgreek}
\usepackage[linkcolor = blue, citecolor = blue, urlcolor = blue, colorlinks = true]{hyperref}
\usepackage{amssymb}
\usepackage{bm}
\usepackage[capitalise]{cleveref}
\usepackage{textcomp}
\usepackage{hyperref}
\usepackage[usenames,dvipsnames]{xcolor}
\usepackage[normalem]{ulem}
\usepackage{wasysym,siunitx}

\usepackage{siunitx}

\newcommand{\Ecoli}{{\it E.~coli}}

\renewcommand{\imath}[0]{\mathsf{i}}

\hypersetup{colorlinks=true, linkcolor=blue!50!black, urlcolor=blue!50!black, citecolor=blue!50!black}

\definecolor{ABpurple}{RGB}{128, 0, 128}
\definecolor{ABred}{RGB}{255, 0, 0}
\definecolor{ABgreen}{RGB}{0, 255, 0}
\definecolor{ABbrown}{RGB}{128, 64, 0}
\definecolor{ABblue}{RGB}{0, 0, 255}

\begin{document}
\title{Characterization and Control of the Run-and-Tumble Dynamics of \textit{Escherichia Coli}}
\author{Christina Kurzthaler}
\thanks{C.K. and Y.Z. contributed equally.}
\email{ckurzthaler@pks.mpg.de}
\affiliation{Max Planck Institute for the Physics of Complex Systems, 01187 Dresden, Germany}
\affiliation{Center for Systems Biology Dresden, 01307 Dresden, Germany}
\affiliation{Department of Mechanical and Aerospace Engineering, Princeton University, Princeton, New Jersey 08544, USA}
\affiliation{Institut f\"ur Theoretische Physik, Universit\"at Innsbruck, Technikerstra{\ss}e 21A, A-6020 Innsbruck, Austria}
\author{Yongfeng Zhao}
\thanks{C.K. and Y.Z. contributed equally.}
\email{yfzhao2021@suda.edu.cn}
\affiliation{Center for Soft Condensed Matter Physics and Interdisciplinary Research \& School of Physical Science and Technology, Soochow University, Suzhou 215006, China}
\affiliation{School of Biomedical Sciences, Li Ka Shing Faculty of Medicine, University of Hong Kong, Pok
Fu Lam, Hong Kong, PR China}
\affiliation{Universit\'e de Paris, MSC, UMR 7057 CNRS, 75205 Paris, France}
\affiliation{ School of Physics and Astronomy and Institute of Natural Sciences, Shanghai Jiao Tong University, Shanghai 200240, China}
\author{Nan Zhou}
\affiliation{ZJU-Hangzhou Global Scientific and Technological Innovation Center, Zhejiang University, Hangzhou  311200, China}
\author{Jana Schwarz-Linek}
\affiliation{School of Physics and Astronomy, University of Edinburgh, James Clerk Maxwell Building, Peter Guthrie Tait Road, Edinburgh EH9 3FD, United Kingdom}
\author{Clemence Devailly}
\affiliation{School of Physics and Astronomy, University of Edinburgh, James Clerk Maxwell Building, Peter Guthrie Tait Road, Edinburgh EH9 3FD, United Kingdom}
\author{Jochen Arlt}
\affiliation{School of Physics and Astronomy, University of Edinburgh, James Clerk Maxwell Building, Peter Guthrie Tait Road, Edinburgh EH9 3FD, United Kingdom}
\author{Jian-Dong Huang}
\affiliation{School of Biomedical Sciences, Li Ka Shing Faculty of Medicine, University of Hong Kong, Pok Fu Lam, Hong Kong, PR China}
\affiliation{CAS Key Laboratory of Quantitative Engineering Biology, Shenzhen Institute of Synthetic Biology, Shenzhen Institutes of Advanced Technology, Chinese Academy of Sciences, Shenzhen 518055, China}
\author{Wilson C. K. Poon}
\affiliation{School of Physics and Astronomy, University of Edinburgh, James Clerk Maxwell Building, Peter Guthrie Tait Road, Edinburgh EH9 3FD, United Kingdom}
\author{Thomas Franosch}
\affiliation{Institut f\"ur Theoretische Physik, Universit\"at Innsbruck, Technikerstra{\ss}e 21A, A-6020 Innsbruck, Austria}
\author{Julien Tailleur}
\email{julien.tailleur@univ-paris-diderot.fr}
\affiliation{Universit\'e de Paris, MSC, UMR 7057 CNRS, 75205 Paris, France}
\author{Vincent A. Martinez}
\email{vincent.martinez@ed.ac.uk}
\affiliation{School of Physics and Astronomy, University of Edinburgh, James Clerk Maxwell Building, Peter Guthrie Tait Road, Edinburgh EH9 3FD, United Kingdom}

\begin{abstract}
We characterize the full spatiotemporal gait of populations of swimming {\it Escherichia coli} using renewal processes to analyze the measurements of intermediate scattering functions. This allows us to demonstrate quantitatively how the persistence length of an engineered strain can be controlled by a chemical inducer and to report a controlled transition from perpetual tumbling to smooth swimming. 
For wild-type {\it E.~coli}, we measure simultaneously the microscopic motility parameters and the large-scale effective diffusivity, hence quantitatively bridging for the first time small-scale directed swimming and macroscopic diffusion.
\end{abstract}
\maketitle

Swimming is key for many
micro-organisms to survive~\cite{Berg:1972,Berg:2008,Poon:2016,Pohl:2017,Riedel:2005,Friedrich:2008,Merchant:2007,Machemer:1972}. Such `active matter' is necessarily far from thermal 
equilibrium~\cite{Romanczuk:2012,Elgeti:2015,Bechinger:2016,o2022time} and displays peculiar transport properties, which
enable foraging~\cite{benichou2011intermittent} and escaping from
harm~\cite{Wadhams:2004}.  The flagellated bacterium {\it Escherichia
  coli} is a model system for active matter
experiments~\cite{poon2013physics,Poon:2016,Wu:2000,Liu:2011,Curatolo:2020,chen2017weak,Arlt2019,Martinez2020}. Much
is known about its genetics, biochemistry, and
ultrastructure, but relating this knowledge to the emerging phenotype, for instance to predict the three-dimensional (3D) pattern of locomotion (or gait) of a swimming population, remains a challenge.

The bacterium's `run and tumble' (RT) dynamics~\cite{Berg:2008} alternates
between persistent motion along the cell's axis and sudden changes of direction. While the motion of isolated flagella has been studied in detail using \textit{in vitro} single-motor experiments~\cite{cluzel2000ultrasensitive,korobkova2004molecular},
 a quantitative characterization of the full 3D gait of swimming populations of multi-flagellated bacteria has been out of reach so far. This stems, in particular, from the need for measurements over length scales ranging from the order of the short-time runs ($\sim 1-\SI{10}{\micro\meter}$) to far beyond the persistence length ($\gtrsim \SI{e2}{\micro\meter}$). 
Assuming exponentially distributed run and tumble durations,  the RT dynamics is predicted to lead to a large-scale diffusion~\cite{Berg:2008,Lovely:1975,Schnitzer:1993,Cates:2013}, but this claim has seldom been demonstrated experimentally, and the underlying assumptions have recently been questioned~\cite{cluzel2000ultrasensitive,Figueroa:2018}. The accurate characterization of RT dynamics will therefore fill an important gap. 

At the same time, while many aspects of the motility of {\it E.~coli} have been brought under direct experimental control~\cite{galajda2007wall,Berg:2008,sokolov2010swimming,diLeonardo:2010,wensink2012meso,saragosti2011directional,nishiguchi2017long,Pohl:2017,Curatolo:2020,grobas2021swarming,Alirezaeizanjani:2020}, including the ability to regulate its run speed by light~\cite{Frangipane:2018,Arlt:2018}, there is currently only limited scope to fine-tune its overall gait compared to synthetic  swimmers~\cite{Howse:2007,bricard2013emergence,yan2016reconfiguring,bauerle2018self,van2019interrupted,nishiguchi2015mesoscopic} because the bacterium's tumble dynamics is difficult to control independently. The aforementioned lack of good methods to quantify its RT dynamics contributes to these difficulties, which in turn limits the use of {\it E.~coli} as a model organism for fundamental active matter research. 

In this letter, we report a full characterization of the 3D gait of {\it E.~coli}, which enables us to demonstrate how the RT dynamics of an engineered strain can be quantitatively tuned from perpetual tumbling to smooth swimming as the concentration of a chemical inducer is varied. These cells can be used in future work under conditions in which their persistence length can be predicted \textit{a priori} once experimental conditions are specified.

We characterize a bacterium's displacement $\Delta{\bf r}(\tau)$ in Fourier space using the intermediate scattering function (ISF):
\begin{equation}\label{eq:ISFdef}
    f({\bf k},\tau) = \left\langle e^{-i {\bf k}\cdot \Delta{\bf r}(\tau)} \right\rangle\;.
\end{equation}
Analyzing the ISF with a renewal theory then allows us  to extract microscopic dynamical parameters such as the particle speed and the run and tumble durations.  
To measure the ISF, we extend conventional differential dynamic microscopy (DDM)~\cite{Martinez:2012} to collect data encompassing both short-length-scale directed-swimming and large-length-scale diffusive regimes. DDM allows us to work in bulk fluid to minimize hydrodynamic interactions with surfaces~\cite{Berg:1990, DiLuzio:2005,Lauga:2006,DiLeonardo:2011,Molaei2014}. It also circumvents the need of single-cell tracking, which requires customized Lagrangian~\cite{Berg:1972,Berg:2008,Liu:2014,Darnige:2017} or
holographic~\cite{Sheng:2006,Bianchi:2017} microscopy and is limited by the need for low cell concentration, statistical accuracy, and short trajectories.  Our data confirm to order of magnitude previous measurements of \Ecoli\/ motility~\cite{Berg:1972}, albeit with a significantly larger run time. We find large length-scale diffusive behavior and compare the extracted diffusivity, $D_{\rm eff}$, with a theoretical
prediction based on the microscopic motility parameters. The predicted $D_{\rm eff}$ is robust against experimental complexities, but 
speed fluctuations contribute $\sim 10\%$ of its value. Below, we focus on the biophysical implications
of our results, and detail our method elsewhere~\cite{Zhao:2021}.

\begin{figure}[t]
\centering
\includegraphics[width = \linewidth, keepaspectratio]{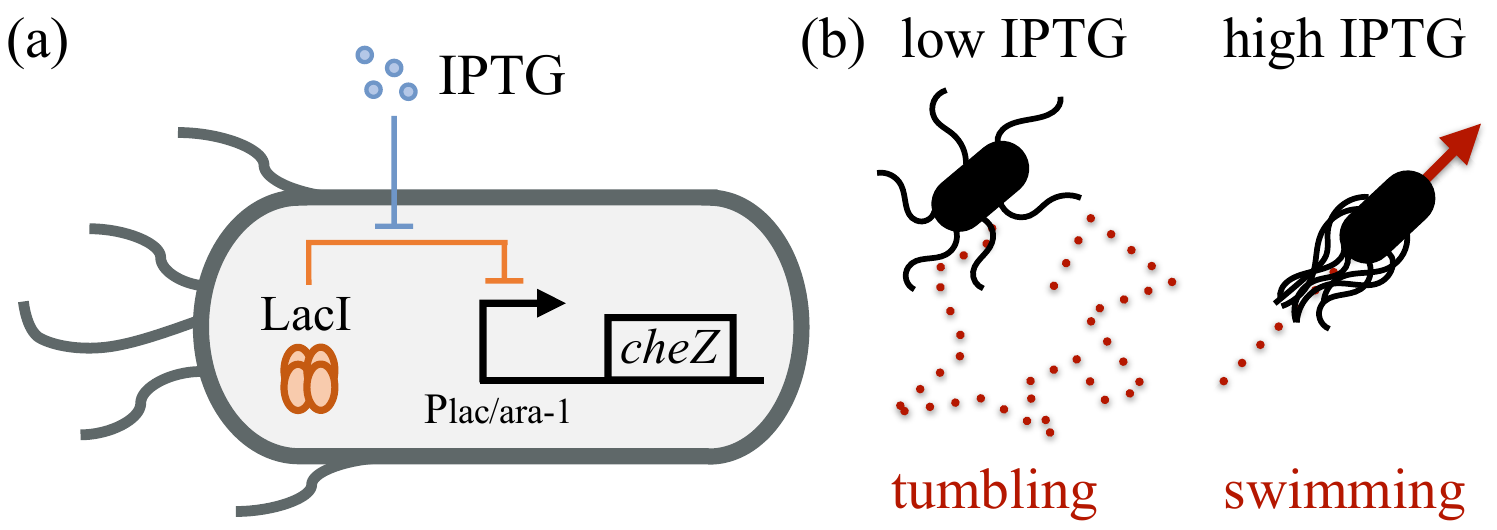}
\caption{Engineered strain NZ1. {\bf (a)} Scheme of the regulation:
  \textit{cheZ} expression driven by Plac/ara-1 is suppressed by the
  LacI suppressor. Exogenously adding IPTG induces \textit{cheZ}
  expression by reducing LacI suppression. {\bf (b)} Cells are expected to
  tumble continuously at low IPTG concentration and to enter a 
  smooth swimming state at high IPTG concentration.}
\label{fig:mutants_map}
\end{figure}

\begin{figure}[tp]
\centering
\includegraphics[width = .9\linewidth, keepaspectratio]{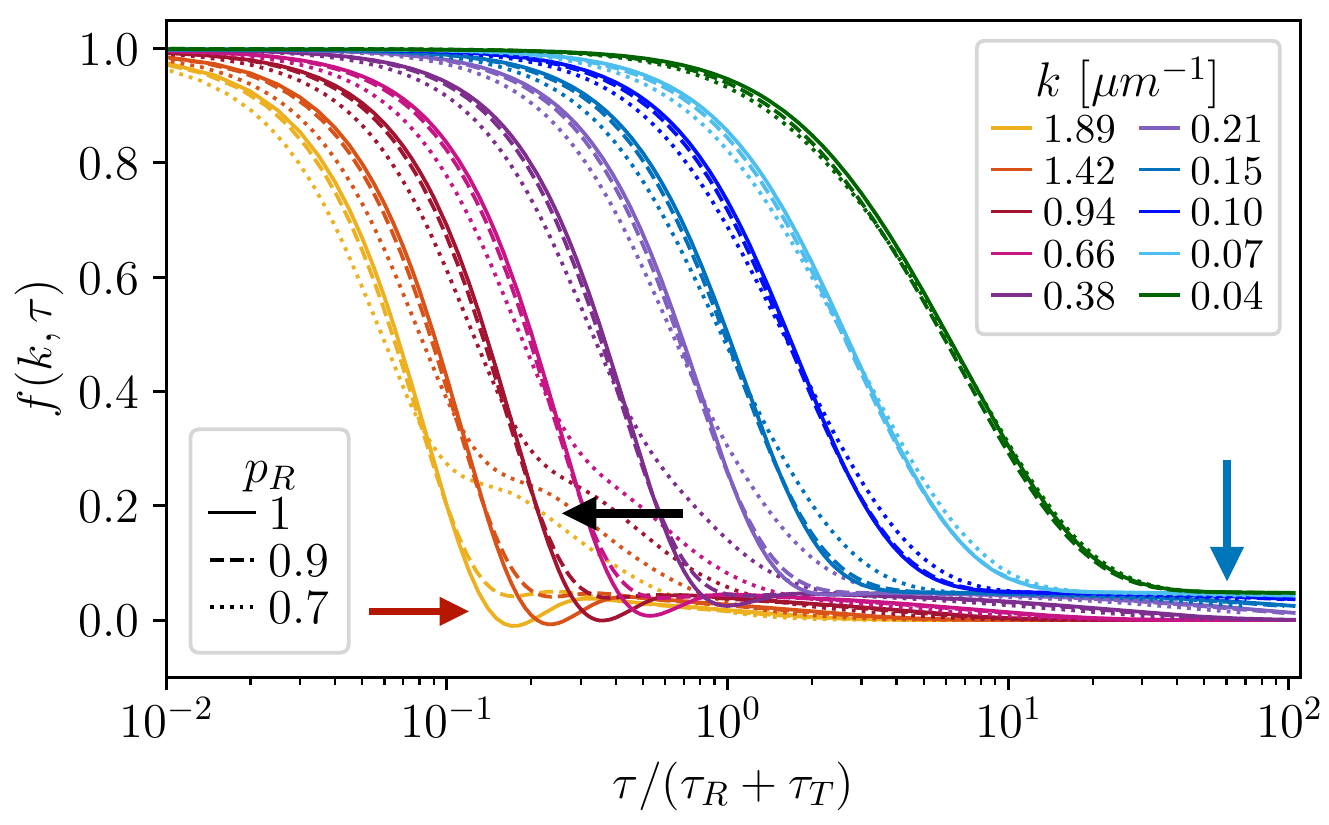}
\caption{Theoretical ISFs for a suspension comprising a
  fraction $\alpha$ of RT bacteria and $1-\alpha$ of diffusing
  cells, for several wavenumbers $k$. We consider different fractions of run time
  $p_R=\tau_R/(\tau_R+\tau_T)$, using $\tau_R=\SI{1}{\second}$ and
  $\tau_T = 0, 0.1, \SI{0.5}{\second}$, with $\bar v =
  \SI{15}{\micro\meter\per\second}$,
  $\sigma_v=\SI{4.5}{\micro\meter\per\second}$,
  $D=\SI{0.3}{\micro\meter\squared\per\second}$, and $\alpha=0.95$. For smooth swimmers
  with fixed speed~$v_0$, $f(k,\tau)=\mathrm{sinc}( v_0
  k\tau)\exp(-D k^2 \tau)$. When $p_R=1$, this leads to clear
  oscillations of $f(k,\tau)$ at small length scales (red arrow). As the
  tumbling rate $\tau_T^{-1}$ increases  (black
  arrow), these oscillations are
  smeared out and the diffusive dynamics of the tumbling bacteria
  eventually leads to a diffusive plateau around $f\sim 1-p_R$. At large but finite times, depending on the value of $k$,
  the diffusive cells may not have decorrelated, leading to a non-zero
  plateau for $f(k,\tau)$ (blue arrow).  }
\label{fig:theory}
\end{figure}

\textit{Bacterial strain.}
We engineered the NZ1 strain by deleting the \textit{cheZ} gene in
\Ecoli\/ K12 and adding the inducible plasmid
Plac/ara-1-cheZ~\cite{Liu:2011} (Fig.~\ref{fig:mutants_map}a).
Deleting \textit{cheZ} suppresses the transition from clock-wise
 to counter-clock-wise flagella rotation, so that cells
tumble permanently.  The plasmid restores expression of \textit{cheZ} at a rate dependant on the concentration of
Isopropyl $\beta$-d-1-thiogalactopyranoside (IPTG). It is expected,
though never yet confirmed, that tuning the concentration of IPTG during the growth of this strain allows the control of
 RT dynamics (Fig.~\ref{fig:mutants_map}b). Bacteria were cultured and re-suspended carefully 
in motility buffer~\cite{Poon:2016} to ensure a very high
final motile fraction, $\alpha\gtrsim95\%$~\cite{supp}.

\begin{figure*}[tp]
\centering
\includegraphics[width = \linewidth, keepaspectratio]{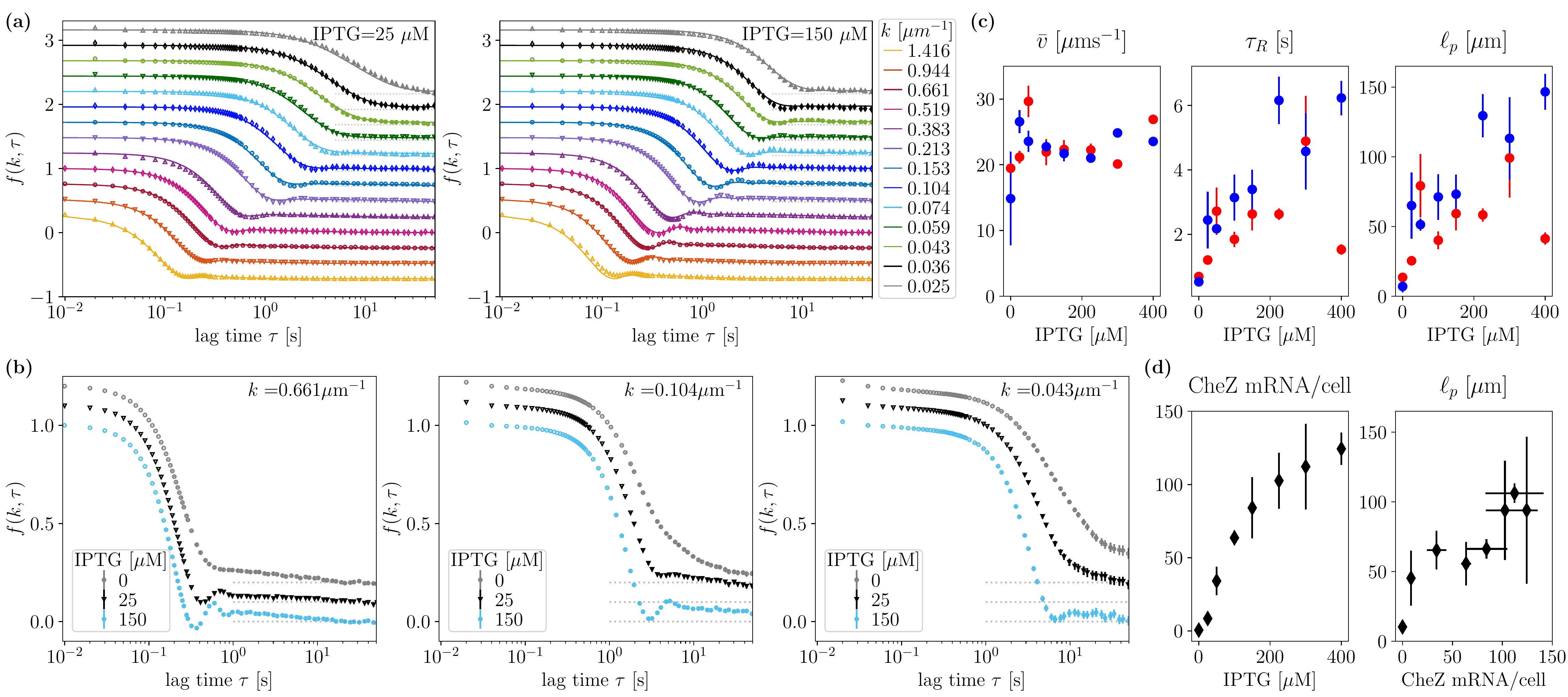}
\caption{Engineered \emph{E. coli} strain NZ1. {\bf (a)} ISFs for IPTG
  concentrations $25\, \upmu$M and $150\, \upmu$M. The
  ISFs are shifted vertically and gray dotted lines correspond to
  $f=0$. Symbols and lines represent experiments and fits to the
  theory, respectively, and different colors correspond to different wavenumbers $k$.  {\bf (b)} Comparison of the ISFs for
  different IPTG concentrations and wavenumbers. {\bf (c)} Average speed $\bar v$, run time
  $\tau_R$, and persistence length $\ell_p$, as a function of IPTG
  concentration. Red and blue symbols correspond to two biological
  replicates. The error is estimated from successive measurements of
  the ISF using the same sample. {\bf (d)} Independent
  measurements of the number of CheZ mRNA per cell at various IPTG
  concentrations ({left panel}) allow correlating the persistence length $\ell_p$ and the 
  number of CheZ mRNA per cell (right panel).
\label{fig:mutants_ISF}}
\end{figure*} 

\textit{ISF measurement and analysis.}  To characterize the gait of the
NZ1 strain at different IPTG concentrations, we first measured its
ISF by DDM. We then fitted it to the calculated ISF of a well-established model
of RT
bacteria~\cite{Berg:1972,Schnitzer:1993,celani2010bacterial,chatterjee2011chemotaxis,Angelani:2013} that is modified to account for recently-observed 
 intrinsic fluctuations of the propulsion speed~\cite{Turner:2016}. In
this model, bacteria run in quasi straight lines at speed $v$ until
they enter a tumbling phase, at rate $\tau_R^{-1}$, during which they
fully randomize their orientations. They resume swimming at rate
$\tau_T^{-1}$ with a new swim speed, sampled from a Schulz
distribution $p(v)$ characterized by a mean velocity $\bar v$ and a
standard deviation $\sigma_v$~\cite{Martinez:2012}. (For an alternative way to account for swimming-speed fluctuations, see Ref.~\cite{Zhao:2021}.) In addition, cells
diffuse translationally with diffusivity $D$ during both run and tumble
phases.  There is also a fraction $1-\alpha$ of
non-motile cells that undergo Brownian motion only, also with diffusivity
$D$~\footnote{Since bacteria are anisotropic, a single $D$ is a crude surrogate of the short-time diffusivity matrix elements, but it helps to fit experimental data of swimming cells. A full matrix description requires a larger fraction of passive cells and a different model~\cite{Kurzthaler:2016}, and is outside our scope.}. The ISF for a non-interacting \emph{E. coli} suspension predicted by this model reads:
\begin{align}
f(k,\tau)=\alpha f_{RT}(k,\tau)+(1-\alpha)e^{-Dk^2\tau}\;, \label{eq:ISF_theo}
\end{align}
where $f_{RT}(k,\tau)$ is the ISF of RT
bacteria~\cite{Zhao:2021}. Measuring this ISF for a wide range of $k$ and $
\tau$ values then allows disentangling the
contributions of diffusion, swimming, and tumbling to the dynamics (Fig.~\ref{fig:theory}).  Fitting Eq.~\eqref{eq:ISF_theo} to data finally yields the set of kinetic parameters $\{\bar v, \sigma_v, \tau_R, \tau_T, \alpha,D\}$.

To measure the ISF experimentally, we imaged cells in sealed capillaries on a fully-automated
inverted bright-field microscope with a sCMOS camera. A full
characterization of RT dynamics requires accessing length scales much
greater than the persistence length $\ell_p$ in all directions. This
necessitated a large depth of field at low $k$ to ensure that
bacteria remain in view over large distances in 3D. We
thus consecutively recorded movies at $2\times$ and $10\times$
magnifications and extracted the corresponding ISFs for $k<
\SI{0.9}{\per\micro\meter}$ and $k\geq \SI{0.9}{\per\micro\meter}$
using DDM~\cite{Wilson:2011,Martinez:2012}, which are then fitted  to our renewal theory using a numerical protocol
described elsewhere~\cite{Zhao:2021}.

\textit{IPTG-induced transition from tumbling to swimming.} We grew suspensions of the NZ1 strain at several IPTG concentrations and measured the ISFs. Representative data over approximately four decades in time and two decades in length are shown in Fig.~\ref{fig:mutants_ISF}a. Oscillations  typical of persistent swimmers (Fig.~\ref{fig:theory}) are seen most clearly at the higher IPTG concentration and high $k$ values. 

In more detail, Fig~\ref{fig:mutants_ISF}b compares the ISFs at a given $k$ for
three IPTG concentrations. In the absence of IPTG, diffusion dominates
and the oscillations are absent. At
IPTG=25$\upmu$M, oscillations are seen for
$k=0.66\upmu\text{m}^{-1}$ and $k=0.1\upmu\text{m}^{-1}$ (corresponding to length scales of $\sim 2\pi/k \approx 10$ and \SI{60}{\micro\meter} respectively), while data at the
smallest $k$ shows a smooth decay: the RT dynamics becomes effectively
diffusive on such a large scale. At the highest IPTG concentration,
150$\upmu\text{M}$, oscillations are seen at all scales, showing a
strong enhancement of the persistence length. To our knowledge, this is the first demonstration of the quantitative tuning of the 3D gait of {\it E.~coli} by varying external conditions. 

Our protocol also allows us to quantify phenotypic heterogeneity.  We repeated such measurements for 8 IPTG concentrations, using two
biological replicates and typically 3 to 4 successive measurements of
$f(k,\tau)$ per replicate and IPTG concentration. The fitted 
kinetic parameters are then averaged for each replicate and plotted in
Fig.~\ref{fig:mutants_ISF}c. The small error bars show that successive measurements of
$f(k,\tau)$ at each condition and for each biological replicate yielded consistent results. The observed variability between replicates (compare red and blue data points) therefore quantifies the degree of phenotypic heterogeneity in a clonal population.

Increasing IPTG leads to a
robust increase of the persistence length, $\ell_p=\bar v \tau_R$, which translates into oscillations in the ISF,
Figs.~\ref{fig:mutants_ISF}a-b. Capturing quantitatively such
fine-grained features requires very good statistics, a narrow speed
distribution, and a low fraction of non-motile cells, a challenge that is met by the experimental protocol described in~\cite{supp}.  Since there is little variability in the average speed ($\bar v \simeq \SI{23.5}{\micro\meter\per\second}$ at all finite IPTG
concentrations), the two orders of magnitude increase in $\ell_p$ is the result of a dramatic increase in the run duration. 

Given the high speed of our NZ1 strain and the long run
durations at high IPTG concentrations, the extraction of
$\tau_R$ requires sampling large length scales, at which we find that the tumbling times $\tau_T$
cannot be reliably measured. We have tested the robustness of our
results with respect to $\tau_T$ by fixing it at $\tau_T=0.1,0.2,0.3$~s and extracting the remaining kinetic parameters. We find that the average velocity of the bacteria remains unaffected and the systematic increase of the persistence length persists (Fig.~S1~\cite{supp}). Finally, measurements of the evolution of single motor statistics of WT $E.~coli$ in different environments have been reported before~\cite{korobkova2004molecular}, Fig.~\ref{fig:mutants_ISF}d shows for the first time the measured correlation between $\ell_p$ and the number of CheZ mRNA in the cell. 

\begin{figure}[tp]
\centering
\includegraphics[width = \linewidth, keepaspectratio]{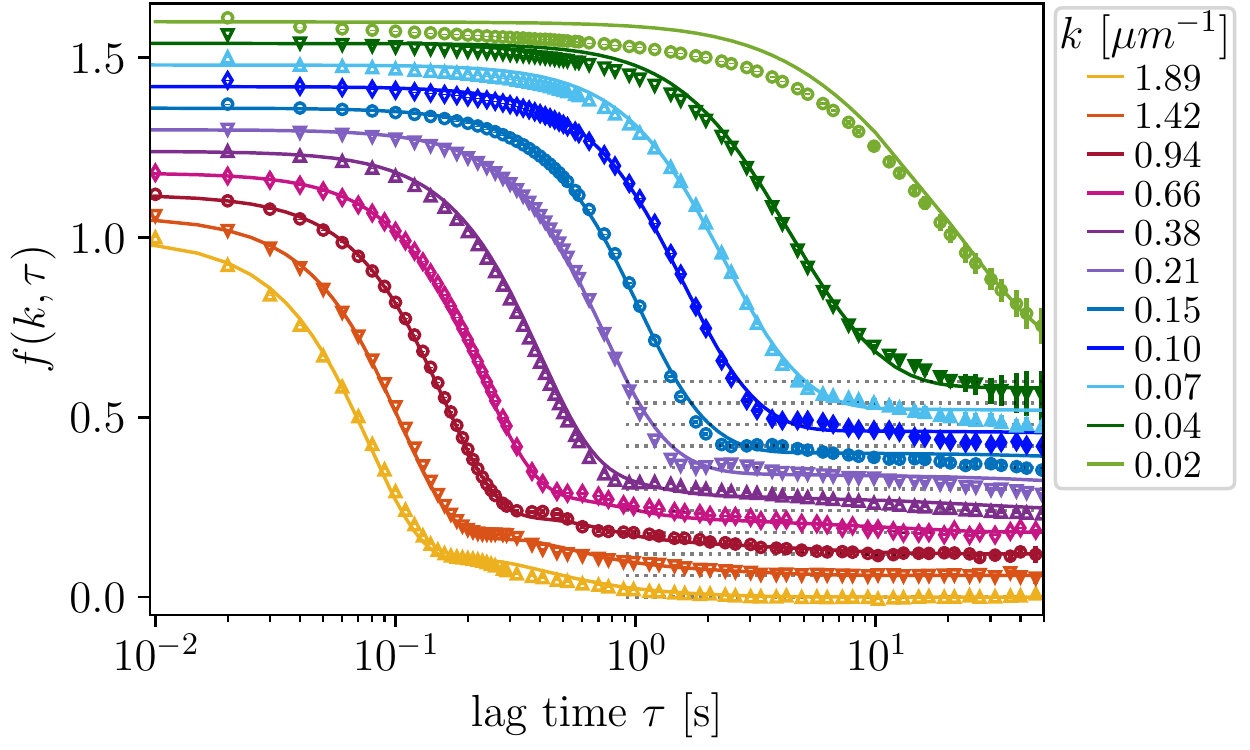}
\caption{Measurements of the ISFs for a WT \textit{E.~coli} suspension 
  (symbols). Fits to Eq.~\eqref{eq:ISF_theo} (lines) lead to
  $\bar v = \SI{15.95}{\micro\meter\per\second}$, $\sigma_v
  = \SI{5.78}{\micro\meter\per\second}$, $D=
  \SI{0.24}{\micro\meter\squared\per\second}$, $\alpha = 0.96$,
  $\tau_R = 2.39 \,\text{s}$, and $\tau_T = 0.38 \,\text{s}$.  The ISFs are shifted vertically as in
  Fig.~\ref{fig:mutants_ISF}a.}
\label{fig:experiments}
\end{figure}

\if{
\begin{figure}[tp]
\centering
\includegraphics[width = \linewidth, keepaspectratio]{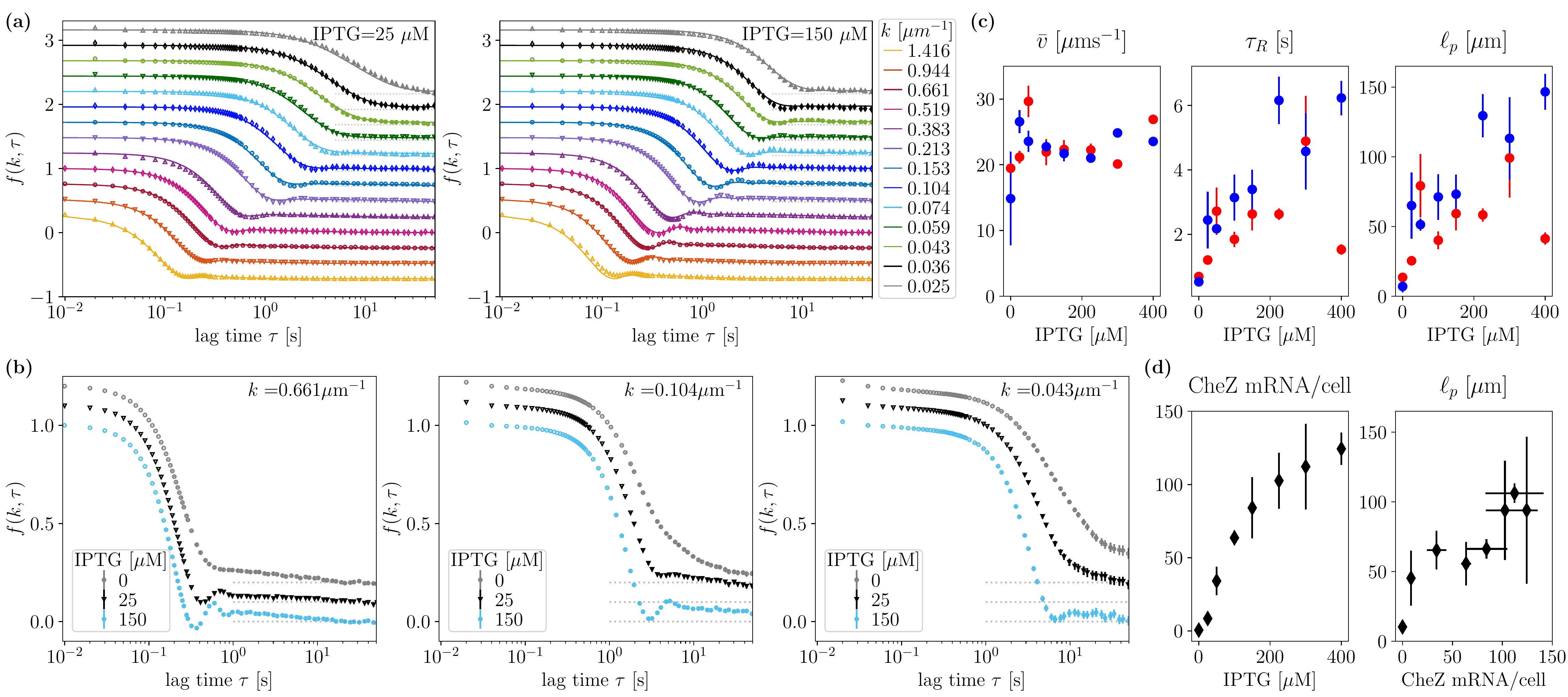}
\caption{Engineered \emph{E. coli} strain (NZ1). (a) Persistence length $\ell_p=\bar v \tau_R$, (b) average swim speed $\bar v \rangle$, and (c) run duration $\tau_R$ extracted from experiments with different IPTG concentrations for two different data sets. To extract the motility parameters, measured ISFs were fitted simultaneously for $k\in[0.025,0.38]\,\si{\per\micro\meter}$. (d) CheZ mRNA/cell as a function of the IPTG concentration in the liquid medium. (e-f) ISFs for $25\, \upmu$M and $150\, \upmu$M IPTG, respectively. The ISFs are shifted with respect to the $y$-axis and the gray dotted lines indicate $y=0$.
\label{fig:mutants}}
\end{figure} }\fi

\textit{Wild-type \textit{E. coli}.}
Figure~\ref{fig:experiments} shows the measured ISFs for a dilute suspension of WT \Ecoli\/ over approximately four decades in time and two decades in length. The ISFs display an intermediate plateau at large $k$, which is a signature of the diffusive motion of both the small non-motile fraction ($\lesssim 5\%$) and of the tumbling cells. The plateau disappears at low $k$ and long times, which reflects the randomization of the swimming direction. 

The WT data are well fitted by our renewal theory~\cite{Zhao:2021} at all $k$ (Fig.~\ref{fig:experiments}, solid lines). We find that $96\pm 0.1\%$ of the bacteria swim at a mean speed 
$\bar v = 16\pm 0.1 \,\si{\micro\meter\per\second}$ with standard deviation $\sigma_v = 5.78\pm 0.13 \,\si{\micro\meter\per\second}$ (errors are obtained by a Jackknife resampling method~\cite{Efron:1981}). 

The lower WT speed (cf.~$\bar v \approx \SI{24}{\micro\meter\per\second}$ for NZ1) allows us to fit the mean run and tumble durations: $\tau_R = 2.39\pm 0.11\,\si{\second}$ (so that $\ell_p=\bar v \tau_R= 38\pm2\,\si{\micro\meter}$) and $\tau_T=0.38\pm 0.02\,\si{\second}$, giving a run fraction of  $p_R=\tau_R/(\tau_R+\tau_T)\simeq0.86$. Original, and still widely-cited, measurements also gave $p_R\simeq0.86$, but with $\tau_R=0.86\pm 0.20\,$\si{\second} and $\tau_T=0.14\pm0.03\,$\si{\second}~\cite{Berg:1972}. These results were obtained by tracking $35$ bacteria, while we averaged over $\sim10^4$--$10^6$ bacteria. 

\begin{figure}[t]
\centering
\includegraphics[width = 0.9\linewidth, keepaspectratio]{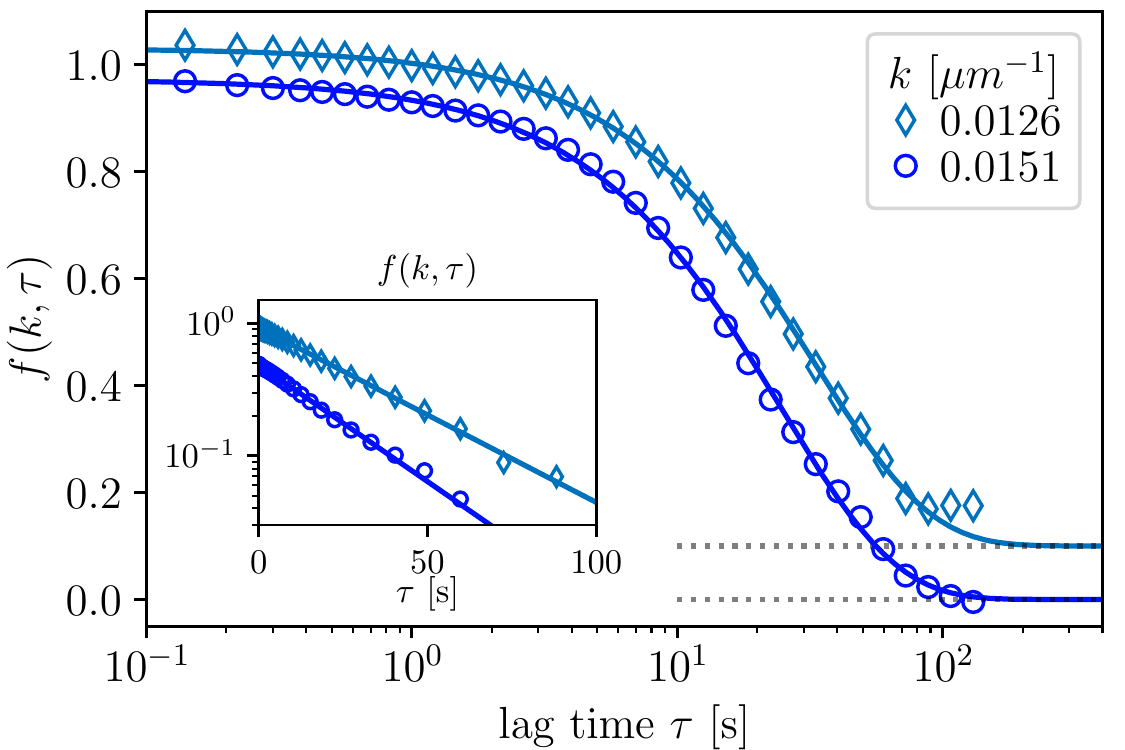}
\caption{Measured ISFs of WT \textit{E.~coli} at
  $k=0.0126\si{\micro\meter}^{-1}$ and
  $k=0.0151\si{\micro\meter}^{-1}$ (symbols) fitted against the ISF of
  diffusive particles (lines), giving effective diffusivities
  $D_{\rm eff}=192\si{\micro\meter\squared\per\second}$ and $D_{\rm
    eff}= 178\si{\micro\meter\squared\per\second}$, respectively. The
  ISFs are shifted vertically as in
  Fig.~\ref{fig:mutants_ISF}a.  }
\label{fig:WT-2k}
\end{figure}

Finally, our data allow us to probe a range of length and time scales large enough to bridge short-scale directed swimming and large-scale diffusive motion. This is an important challenge since recent experiments have questioned the experimental relevance of exponentially-distributed run and tumble durations~\cite{cluzel2000ultrasensitive,Figueroa:2018} \if{(JT: Can we check whether Cluzel mentions Levy flight and what are the exponents? Do they violate CLT? YZ: In Ref. [23] of Cluzel, they mentioned in Fig. 2c that the cumulative distribution from experiments with WT E. coli has a power-law exponent of -1.2, corresponding to a power-law exponent of the run-times distribution of -2.2. The exponent -2.2 was cited by Tu and Grinstein, PRL 94, 208101 (2005). 
The exponent is larger than -3 and CLT is violated.)}\fi. The ISF of purely diffusive particles, $f(k,\tau)=A\exp(-k^2D_\text{eff}\tau)$, where $A$ is a constant, gives a good account of our large-scale data (shown for two values of $k$ in  Fig.~\ref{fig:WT-2k}). Averaging over our two smallest
values of $k$ leads to $\langle D_\text{eff} \rangle=185\pm7\,\si{\micro\meter\squared\per\second}$.
 
In an RT model with exponentially distributed run and tumble durations, $\langle D_{\rm
  eff}^{\rm th}\rangle=({\bar v}^2 + \sigma_v^2) \tau_R^2/(3
\tau_R+3\tau_T)$. Using parameters from fitting the measured ISF (caption, 
Fig.~\ref{fig:experiments}), we find $\langle D_{\rm eff}^{\rm th} \rangle =
198\pm11\,\si{\micro\meter\squared\per\second}$, which is remarkably close to the measured value. 
Interestingly, the $\sigma_v^2$ term arising from velocity fluctuations contributes to $\sim 10\%$
of the value of $\langle D_{\rm eff}^{\rm th}\rangle $. Note that our measured $D_{\rm eff}$ is three orders of magnitude larger than the fitted single-particle diffusivity, {$D = 0.24\pm 0.01 \,\si{\micro\meter\squared\per\second}$}, which highlights the ability of our protocol to provide information on active-particle dynamics over a large spatiotemporal range. 

{\it Conclusion.---} By characterizing the dynamics of {\it E.~coli} over a wide range of length and time scales, we demonstrated for the first time how the tumbling dynamics of an engineered strain can be quantitatively tuned independently of the swimming speed. 
Furthermore, we have characterized to a high statistical accuracy the full 3D gait of WT \textit{E.~coli} in bulk suspensions, from small-scale persistent motion to large-scale diffusion. We have shown that a microscopic RT model with exponentially distributed RT durations describes both regimes. The use of more realistic distributions~\cite{Figueroa:2018} can be accommodated in our approach, but is unlikely to change significantly any of our conclusions.

Our work lays the foundation for the high-throughput study of the swimming
gait of a variety of microorganisms, such as the RT
pattern of \emph{Bacillus subtilis}~\cite{Turner:2016} or the
run-reverse motion of marine bacteria~\cite{Johansen:2002} and
archaea~\cite{Thornton:2020}, using standard microscopy. More
generally, the ability of microorganisms to respond to chemical
gradients, i.e., chemotaxis, is a vital part of their foraging and
survival strategy. Our method in combination with
spatiotemporally-resolved DDM \cite{Arlt2019} opens the way to the
high-throughput study of such response at the population level.

\begin{acknowledgments}
{\it Acknowledgments.--} This work was supported by the Austrian
Science Fund (FWF) via  P35580-N and the Erwin Schr{\"o}dinger
fellowship (J4321-N27), the European Research Council Grant
AdG-340877-PHYSAPS, the ANR grant Bactterns, the Shenzhen Peacock Team
Project (KQTD2015033117210153), and the National Key Research and
Development Program of China (2021YFA0910700).
\end{acknowledgments}
\vspace{-0.5cm}
\bibliography{literature}

\end{document}


\title{Characterization and Control of the Run-and-Tumble Dynamics of \textit{Escherichia Coli}: Supplemental Material}
\author{Christina Kurzthaler}
\thanks{C.K. and Y.Z. contributed equally.}
\email{ckurzthaler@pks.mpg.de}
\affiliation{Max Planck Institute for the Physics of Complex Systems, 01187 Dresden, Germany}
\affiliation{Center for Systems Biology Dresden, 01307 Dresden, Germany}
\affiliation{Department of Mechanical and Aerospace Engineering, Princeton University, Princeton, New Jersey 08544, USA}
\affiliation{Institut f\"ur Theoretische Physik, Universit\"at Innsbruck, Technikerstra{\ss}e 21A, A-6020 Innsbruck, Austria}
\author{Yongfeng Zhao}
\thanks{C.K. and Y.Z. contributed equally.}
\email{yfzhao2021@suda.edu.cn}
\affiliation{Center for Soft Condensed Matter Physics and Interdisciplinary Research \& School of Physical Science and Technology, Soochow University, Suzhou 215006, China}
\affiliation{School of Biomedical Sciences, Li Ka Shing Faculty of Medicine, University of Hong Kong, Pok
Fu Lam, Hong Kong, PR China}
\affiliation{Universit\'e de Paris, MSC, UMR 7057 CNRS, 75205 Paris, France}
\affiliation{ School of Physics and Astronomy and Institute of Natural Sciences, Shanghai Jiao Tong University, Shanghai 200240, China}
\author{Nan Zhou}
\affiliation{ZJU-Hangzhou Global Scientific and Technological Innovation Center, Zhejiang University, Hangzhou 311215, China}
\author{Jana Schwarz-Linek}
\affiliation{School of Physics and Astronomy, University of Edinburgh, James Clerk Maxwell Building, Peter Guthrie Tait Road, Edinburgh EH9 3FD, United Kingdom}
\author{Clemence Devailly}
\affiliation{School of Physics and Astronomy, University of Edinburgh, James Clerk Maxwell Building, Peter Guthrie Tait Road, Edinburgh EH9 3FD, United Kingdom}
\author{Jochen Arlt}
\affiliation{School of Physics and Astronomy, University of Edinburgh, James Clerk Maxwell Building, Peter Guthrie Tait Road, Edinburgh EH9 3FD, United Kingdom}
\author{Jian-Dong Huang}
\affiliation{School of Biomedical Sciences, Li Ka Shing Faculty of Medicine, University of Hong Kong, Pok Fu Lam, Hong Kong, PR China}
\affiliation{CAS Key Laboratory of Quantitative Engineering Biology, Shenzhen Institute of Synthetic Biology, Shenzhen Institutes of Advanced Technology, Chinese Academy of Sciences, Shenzhen 518055, China}
\author{Wilson C. K. Poon}
\affiliation{School of Physics and Astronomy, University of Edinburgh, James Clerk Maxwell Building, Peter Guthrie Tait Road, Edinburgh EH9 3FD, United Kingdom}
\author{Thomas Franosch}
\affiliation{Institut f\"ur Theoretische Physik, Universit\"at Innsbruck, Technikerstra{\ss}e 21A, A-6020 Innsbruck, Austria}
\author{Julien Tailleur}
\email{julien.tailleur@univ-paris-diderot.fr}
\affiliation{Universit\'e de Paris, MSC, UMR 7057 CNRS, 75205 Paris, France}
\author{Vincent A. Martinez}
\email{vincent.martinez@ed.ac.uk}
\affiliation{School of Physics and Astronomy, University of Edinburgh, James Clerk Maxwell Building, Peter Guthrie Tait Road, Edinburgh EH9 3FD, United Kingdom}

\date{\today}

\maketitle
\setcounter{equation}{0}
\renewcommand\theequation{S.\arabic{equation}}
\renewcommand\thefigure{S.\arabic{figure}}


Most of the experimental, imaging, and analysis protocols are either
standard or can be found in previous
publications~\cite{Martinez:2012,Poon:2016,Wilson:2011,Liu:2011,Cerbino:2008,Kurzthaler:2018:janus}. To
make the reading as self-content as possible, we nevertheless briefly
describe them below, as well as specify the culture conditions and
biological protocols used in our study. The numerical protocol and
its validation are presented in a joint submission~\cite{zhao:2021}.

\section{Experimental protocol, imaging, and DDM~\label{sec:experiments}}
\subsection{Samples} We used two strains of \Ecoli\/ K12: a wild-type AB1157, which have been described elsewhere \cite{Poon:2016}, and the engineered strain (NZ1) obtained from AMB1655.  We cultured both strains using standard protocols described elsewhere \cite{Poon:2016}. Briefly, an overnight culture was obtained by inoculating a single colony into 10 mL of Luria broth followed by incubation at \SI{30}{\degreeC}/$200\,$rpm for 16 to \SI{18}{\hour}. Then this was inoculated into \SI{35}{\milli\liter} of Tryptone broth (TB) medium (1:100 dilution), which was incubated for \SI{4}{\hour} (\SI{30}{\degreeC}/$200\,$rpm) to obtain a late-exponential-phase culture. Cells were then harvested and concentrated by gentle filtration (\SI{0.45}{\micro\meter} Immobilon filters, Millipore) and resuspended in motility buffer (pH 7.0, 6.2 mM K2HPO4, 3.8 mM KH2PO4, 67 mM NaCl, and 0.1 mM). Here, we used a single filtration step to avoid damage to the flagella, which yields suspensions (2-4~\si{\milli\liter} at $\approx 5\times 10^8$cells/ml) with higher motile fraction ($> 95\%$) and a narrower swimming speed distribution than for multiple filtration steps~\cite{Poon:2016}. For the engineered strain NZ1, we added IPTG at a given concentration in the range (0-$400\uM$) in TB.

\subsection{Imaging} Results from computer simulations~\cite{zhao:2021} suggested one needs to access a wide range of wave numbers $k$ corresponding to length scales $\ell$ ranging from the bacterial size to over the run length (i.e. $\sim \SI{100}{\micro\meter}$) for a reliable analysis. Thus, a large depth of field is required at low $k$ values. We consecutively recorded movies ($512\,$pxl $\times 512\,$pxl) using two magnifications ($10\times$/0.3, 8000 images at $100\,$fps and $2\times$/0.06, 32000 images at $50\,$fps), an inverted bright-field microscope (Nikon Ti E), and a sCMOS camera (Hamamatsu Orca Flash 4.0 v2, using $2 \times 2$ binning). We restricted  the condenser aperture to $\approx \SI{5}{\milli\meter}$ diameter so that both $10\times$ and $2\times$ imaging could be performed using the same illumination conditions, which yields a depth of field of about \SI{300}{\micro\meter} for the lowest magnification. Due to the large $\ell$ (in 3D) required for reliable analysis, we used glass capillaries with a large height (\SI{1}{\milli\meter}) and focused in the middle. Note that the depth of field for $10\times$ objective is too small for the required large length-scales. Finally, the sampling volumes corresponding to high and low magnifications were estimated to $\approx 10^{-5}$ and $3\times 10^{-3}\si{\milli\liter}$ (assuming a depth of field of $\approx 15$ and \SI{300}{\micro\meter}, respectively) corresponding to $\approx 10^4$ and $10^6$ cells in the field of view, respectively, for a suspension with $\approx 5\times 10^8$ cells/mL.

The data for the mutants are fitted using $k\in[0.036,1.42]\upmu$m$^{-1}$ and for the data of the wild type we included $k\in[0.04, 1.89]\upmu$m$^{-1}$.

\subsection{Differential Dynamic Microscopy (DDM)} We then analyzed each movie using DDM to extract the ISFs, following~\cite{Cerbino:2008, Martinez:2012}. Briefly, we computed the differential image correlation function (DICF), $g(\vec{k},\tau)$, i.e., the power spectrum of the difference between pairs of images separated by time $\tau$, by $g(\vec{k}, \tau)=\left\langle\left|I(\vec{k}, t+\tau)-I(\vec{k}, t)\right|^2\right\rangle_t$, where $I(\vec{k},t)$ is the Fourier transform of the image $I(\vec{r},t)$ and $\langle\cdot\rangle_t$ denotes an average over time $t$. Under suitable imaging conditions and for isotropic motion, the DICF yields the ISF \cite{Wilson:2011,Martinez:2012,Reufer2012}, $f(k,\tau)$, via
\begin{align}
g(k,\tau) =\left \langle g(\vec{k},\tau)\right\rangle_{\vec{k}}= A(k)\left[1-f(k,\tau)\right]+B(k)
\end{align}
with $\left\langle\cdot\right\rangle_\vec{k}$ denoting average over $\vec{k}$ and $A(k)$ and $B(k)$ the signal amplitude and instrumental noise, respectively. These coefficients are obtained from the plateau of $g(k,\tau)$ at long and short times, where the ISF approaches $f(k,\tau\to \infty)\rightarrow 0$ and $f(k,\tau\to 0)\rightarrow 1$, respectively. 

\section{Engineered \Ecoli \ (NZ1)~\label{sec:mutants}}
\subsection{Fitting results} 
\begin{figure}
    \centering
    \includegraphics[width=\textwidth]{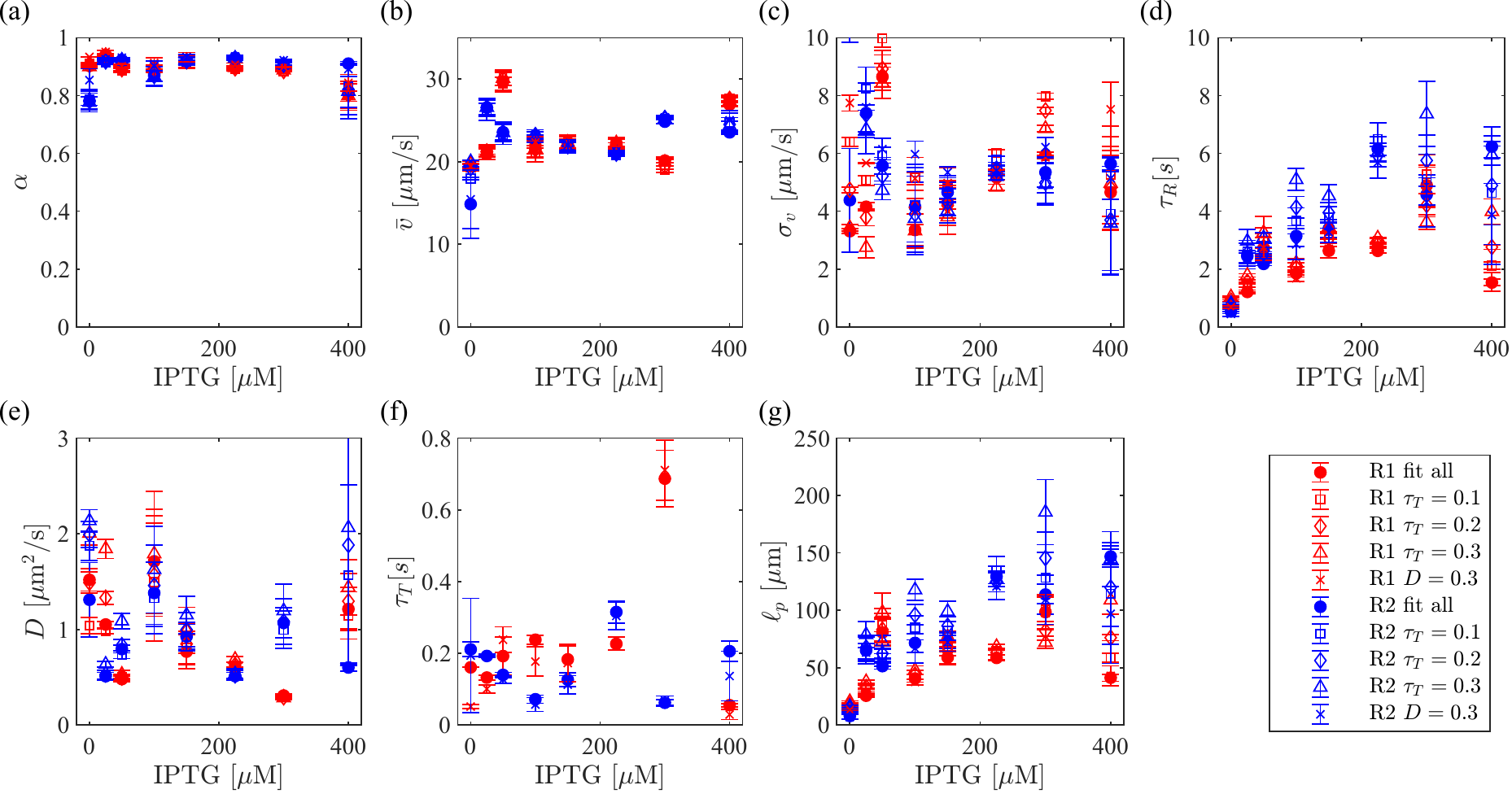}
    \caption{The fitting estimates of the two replicates of NZ1 as a function of IPTG concentration. The estimates with fitting all the parameters are shown as solid symbols. The open symbols show the estimates with fixing $\tau_T$ at 0.1, 0.2, 0.3 s, or $D$ at 0.3 $\si{\square\micro\meter\per\second}$.}
    \label{fig:fitting_estimates}
\end{figure}

The fitting estimates are shown in Fig. \ref{fig:fitting_estimates} by the solid symbols. To test the robustness of our results under the fluctuations in the estimates of $D$ and $\tau_T$, we fix $\tau_T$ at 0.1, 0.2, 0.3 s, and $D$ at 0.3 $\si{\square\micro\meter\per\second}$ and extract the remaining parameters. We find that the fraction of moving bacteria, the average velocity, and the average run time remain unaffected. The systematic increase of the persistence length persists. It suggests that the Brownian diffusion constant $D$ cannot be reliably measured in the sample with few non-motile cells. The difficulty in estimating $\tau_T$ is highlighted as well.

\subsection{Estimation of CheZ mRNA per cell} We measure the number of CheZ mRNA per cell by reverse transcription of quantitative PCR as previously described [9]. Briefly, Around 0.5 OD NZ1 cells induced at different concentrations of IPTG in TB was collected by centrifugation. Cell lysates were prepared and spiked with a known amount (around $10$~ng) of external RNA. Total RNA was extracted using the RNeasy Mini Kit (QIAGEN) and $700$~ng of the extracted RNA was treated by DNase I, Amp Grade (Invitrogen) to eliminate genomic DNA and plasmid contamination. Purified RNA was used for reverse transcription using PrimeScriptTM RT reagent Kit (TaKaRa) and the reverse transcribed DNA was then quantified by qPCR using qTOWER3 (Analytik Jena). The copy numbers of CheZ mRNA and the external RNA were calculated according to a standard curve of known quantities of the corresponding linearized plasmid. To estimate the CheZ mRNA copy numbers per cell, a recovery rate, which describes the ratio between the measured and initial amounts of mRNA, was calculated by dividing the measured output by the input amount of the spike mRNA. Then the measured amounts of CheZ mRNA was divided by the recovery rate to obtain its initial amounts per sample. This number was further divided by the number of cells present in the initial sample to calculate the number of CheZ mRNA copies per cell. The number of cells per sample was obtained from the measured absorbance at $600$~nm (OD600). To correlate OD600 to the number of bacteria, the measured absorbance was plotted against the colony forming units (CFU) of NZ1, which approximately gives $1$ OD600 of NZ1$=4.73\cdot108$~CFU. The CheZ mRNA is shown as a function of IPTG in Fig.~4(d) in the main text.

\bibliography{literature}